\documentclass[aps,prper,showpacs,twocolumn,floats,superscriptaddress,10pt]{revtex4-1}
\usepackage{graphicx}
\usepackage{booktabs}
\usepackage{dcolumn}
\usepackage{bm}
\usepackage{amssymb}
\usepackage{nicefrac}

\usepackage[latin3]{inputenc}
\usepackage[makeroom]{cancel}
\usepackage{amsmath}
\usepackage{amsfonts}
\usepackage{amssymb}
\usepackage{color}
\usepackage[left=2cm,right=2cm,top=2cm,bottom=2cm]{geometry}

\usepackage{graphicx} 
\usepackage{dcolumn}
\usepackage{bm}
\usepackage{simplewick}
\usepackage{array}
\usepackage{appendix}

\RequirePackage[
   hyperindex,colorlinks,bookmarksnumbered,
   plainpages=true,pdfstartview=FitH]{hyperref}
\hypersetup{linkcolor=blue,urlcolor=blue,citecolor=blue}
\usepackage{hyperref}

\definecolor{purple}{rgb}{0.5,0,0.6}

\usepackage{ulem}
\begin{document}
\title{Generalized Wiedemann-Franz law in a two-site charge Kondo circuit:\\ 
Lorenz ratio as a manifestation of the orthogonality catastrophe}
\author{M. N. Kiselev}
\affiliation{The Abdus Salam International Centre for Theoretical Physics,  Strada
Costiera 11, I-34151, Trieste, Italy }
\date{\today}
\begin{abstract}
We show that the transport integrals of  the two-site charge Kondo circuits
connecting various multi-channel Kondo simulators satisfy the generalized Wiedemann-Franz law with the universal Lorenz ratios all greater than one. The magic Lorenz ratios are directly related to the Anderson's orthogonality catastrophe in quantum simulators providing some additional universal measure for the strong electron-electron correlations.  We present a full fledged theory for the magic Lorenz ratios
and discuss possible routes for the experimental verifications of the theory.
\end{abstract}
\maketitle
\begin{figure}[b]
\includegraphics[width=75mm,angle=0]{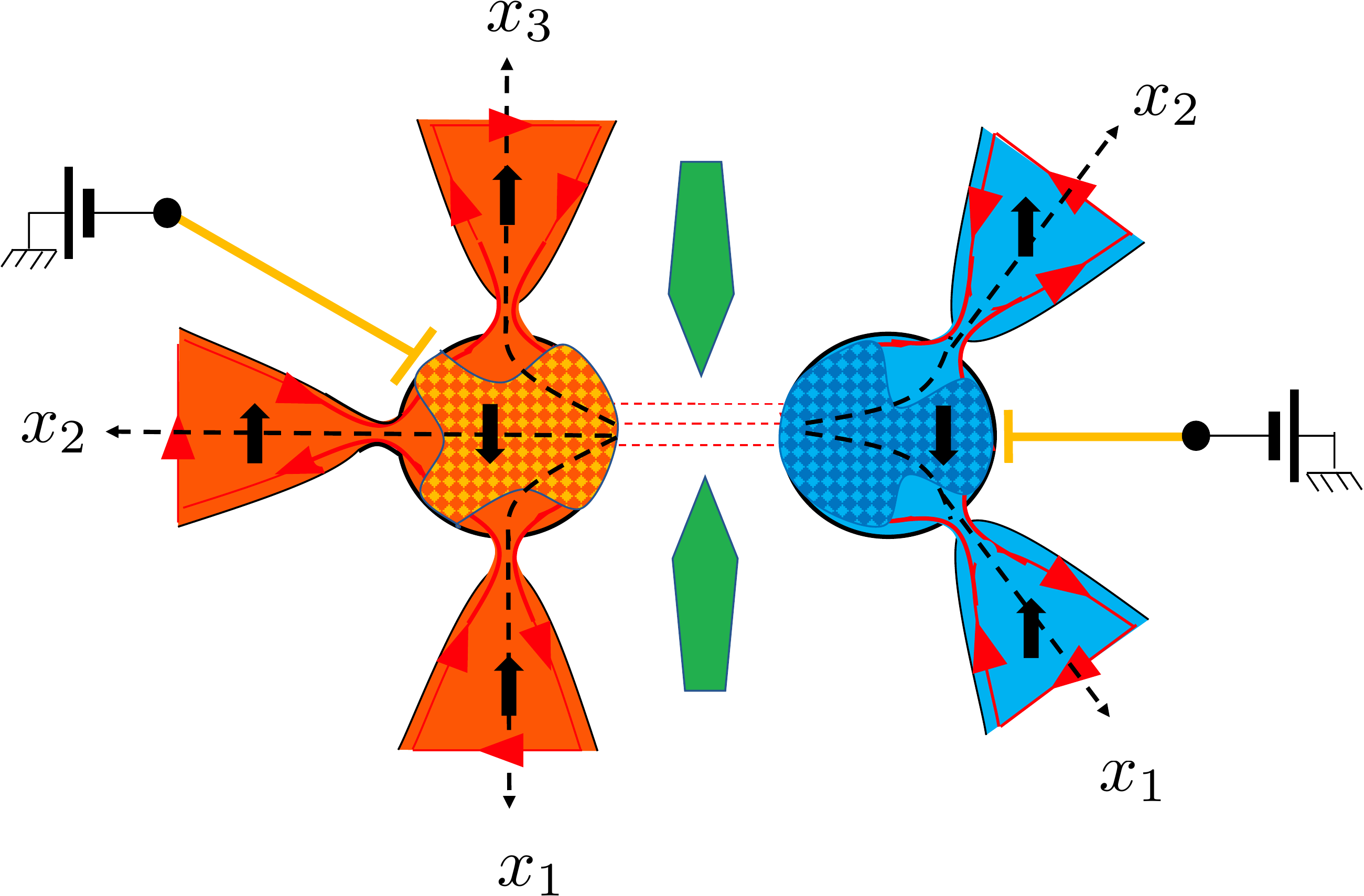}
\caption{Cartoon for the two-site Kondo circuit proposed for the measurement of the charge and heat transport coefficients. Two parts of the circuit fabricated out of 2DEG (orange and blue zones) consisting of Quantum Point Contacts (bottleneck areas)  attached to two Quantum Dots (hatched area inside the circles) are connected through the central tunnelling area(dashed lines).  The yellow plunger gate is used to control a
mesoscopic Coulomb blockade in the QD.  The tunnel contact is adjusted by the gate (green boxes).  For the illustration of the model connecting $N$- and $M$- channels Kondo simulators we show the left (hot, orange) part of the circuit with $N=3$ QPCs  at temperature $T_1$ and the right (cold, blue) part  of the circuit with $M=2$ QPCs being at temperature $T_2<T_1$.  The red lines with arrows along the painted areas denote the Integer Quantum Hall  $\nu=1$ edge states. The bold up/down arrows indicate different pseudospin quantum number outside/inside the quantum dot.  The magic Lorenz ratio (\ref{lora}) for this setup ${\cal R}=15/7$.}
\label{f1} 
\end{figure}

{\color{blue} \textit{Introduction}} - Anderson's orthogonality catastrophe (OC) \cite{orthogonal} describes an effect of a local perturbation on a gas of $N$ fermions.  In order to screen the local impurity potential the quantum many-body states change such a way that a new state becomes orthogonal to the ground state of the system in the 
$N$$\to $$\infty$ thermodynamic limit.
The OC plays an important role in understanding of problems associated with  a sudden (at time $t $$=$$ 0$) excitation of a core electron in an atom (so-called x-ray edge or Mahan's singularity \cite{Mahan,Noz1,Noz2,Noz3}) and Kondo problem \cite{Kondo, Hewson}
as well as many other topics of theoretical and experimental relevance. Typically, OC is manifested in a certain power-law dependences of quantum correlators (response functions) as a function of energy, frequency or temperature  (e.g. local density of states,  x-ray absorption rate etc) directly measured experimentally. Moreover,  the physics behind OC is crucial for a description of many-body systems' dynamics and physics of quantum quenches \cite{speed, OC_UG}. 
In particular, as it was shown recently in \cite{speed}, the dynamics of OC is fully characterized by the quantum speed limit. The ultra-cold atomic gases  Ramsey-interference-type experiments with the impurity atoms \cite{OC_UG} allow one to study the OC in the time domain complementary to the radio-frequency spectroscopy probes of the OC in the frequency domain \cite{OC_UG}.

The OC in strongly correlated condensed matter systems impacts the 
{\it quantum thermodynamic} quantity known as a Wilson ratio (WR) \cite{Hewson}.
The WR for the quantum system  is defined as the ratio of a susceptibility increment 
$\delta \chi/\chi$ to a specific heat increment $\delta C/C$:
\begin{equation}
R_W=\frac{\delta \chi/\chi}{\delta C/C}.
\end{equation}
The WR of the ideal Fermi gas is $R_W$$=$$1$ while e.g. for the quantum impurity single-channel Kondo effect $R_W$$=$$2$ being enhanced by the ratio of the total specific heat to that coming from the spin degrees of freedom \cite{NB}. In general, the the WR depends on the number of the scattered channels and the spin of the impurity \cite{NB} and provides an important measure for the effects of strong electron-electron correlations.

In this {\it Letter} we present some arguments in favour of getting an information
about OC in nano-devices and quantum simulators directly from two {\it quantum transport} low temperature measurements. We argue that the Lorenz ratio $R_L$ 
\begin{equation}
R_L\cdot L_0=\frac{{\cal K}}{G T},
\end{equation}
(here $L_0$$=$$(k_B/e)^2$$\pi^2$$/$$3$ is a Lorenz number, $e$ is the electron's charge and $k_B$ is Boltzmann's constant)
being an universal proportionality coefficient between two {\it quantum transport}  correlation functions: the thermal ${\cal K}$ and electric $G$ conductances can also be used as a measure for the strong interaction effects.  We show that $R_L$ is directly related to the OC physics uniquely characterizing the strongly correlated operational regimes of the quantum simulators.

Two-site Kondo simulators have been theoretically proposed in \cite{thanh2018}
to investigate competing phases associated with Fermi- and Non-Fermi liquid behaviour in different sides of the circuit and its interplay in the quantum charge and heat transport. The idea is to engineer states in a single-site part of the simulator \cite{pierre2, pierre3} by fine-tuning it to a particular regime of a multi-channel charge Kondo effect \cite{Flensberg, Matveev, furusakimatveevprb}.  Finally,  two parts are to be connected  through either a tunnel barrier or a single mode Quantum Point Contact (QPC) to make the circuit operating in different modes of the strongly correlated quantum simulators. 

The single-site charge Kondo circuit is fabricated out of semiconductor heterostructure in an Integer Quantum Hall regime \cite{pierre2, pierre3}. The edge states form a Luttinger Liquid while almost transparent QPCs act as point-like quantum impurities. The large metallic island (Quantum Dot, QD) provided a mesoscopic charge quantization \cite{AleiGlaz}. Adding several QPCs to the circuit is equivalent to creating new Kondo channels. Fine-tuning QD to a special charge degeneracy  point at the Coulomb peaks allowed to treat the two-fold degenerate charge states  as a pseudo-spin and  describe the circuit by the multi-channel Kondo model. The direct manifestation of the two- and three- channel Kondo physics in the single-site Kondo simulators was reported in \cite{pierre2, pierre3}.

The first  experimental realization of the two-site Kondo circuit was done very recently in \cite{two_islands} triggering an immediate interest of the theoretical community. In addition to the quantum critical phenomena being a focal point of \cite{two_islands},
an interesting questions about emerging para-fermions \cite{Karki2022a, Karki2023} characterized by fractional residual entropy and fractional charge have been raised \cite{thanhprl,Karki2022a, Karki2023}. Very recently,  it was suggested to use the charge Kondo simulators for direct observation of the Kondo impurity state and universal screening using charge pseudospin state \cite{Sela1} and also probing single-electron scattering through a non-Fermi liquid charge-Kondo device \cite{Sela2}.  We present some arguments about using quantum heat and charge transport coefficients for shedding a new light on behaviour of Kondo simulators.

{\color{blue} \textit{Model.}} - In this {\it Letter} we consider a two-site Kondo circuit \cite{thanh2018} schematically illustrated by Fig. \ref{f1}. 
The circuit consists of two parts fabricated out of the two-dimensional electron gas, 2DEG, (orange, hot at a temperature $T_1$ and chemical potential $\mu_1$ and light blue, cold
at the temperature $T_2<T_1$  and chemical potential $\mu_2$) connected through a tunnel contact (dashed lines.)  The temperature drop $\Delta T=T_1-T_2$ and the voltage drop $\Delta V=(\mu_1-\mu_2)/e$ occur across the tunnel barrier.  Both parts of the circuit contain QPC (bottleneck areas) and QD (hatched areas inside the big circle).  We assume that the 2DEG is in the Integer Quantum Hall (IQH) regime with $\nu=1$.  The red line denote the edge state.  Each QPC is fine tuned to a low-reflection (high transparency) regime.

The effective model \cite{andreevmatveev, LeHur2002, thanh2010} contains a Gaussian part described by the action $S=\sum_{i=1,2}\left(S^{(i)}_0 + S^{(i)}_C\right)$ (index $i=1$ stands for the left part of the circuit containing $m_1=N$ quantum point contacts (QPC) and $i=2$ used for the right part of the circuit with $m_2=M$ QPCs). The free Euclidean (imaginary time) action
\begin{eqnarray}
S^{(i)}_{0}=\frac{v_{F}}{2\pi}\!\sum_{\alpha=1}^{m_i}\int_{0}^{\beta}\!dt\int_{-\infty}^{\infty}\!dx\!\left[\frac{\left(\partial_{t}\phi_{\alpha}\right)^{2}}{v_{F}^{2}}+\left(\partial_{x}\phi_{\alpha}\right)^{2}\right],\;\;\;\;\;\;
\label{act0}
\end{eqnarray}
represents the bosonized non-interacting fermions \cite{GNTBook} ($\phi_\alpha(x,t)$ are bosonic fields) in the constriction $\alpha=1,.. ,m_i$, $v_F$ is a Fermi velocity, $\beta=1/T$ is an inverse temperature (we adopt the notations $\hbar$$=$$k_B$$=$$e$$=$$1$) \cite{comment1}.   The action $S^{(i)}_C$
\begin{eqnarray}
S^{(i)}_{C}\left(\tau\right)=\int_{0}^{\beta}\!\!dtE^{(i)}_{C}\left[{\color{black} n_\tau(t)}+\frac{1}{\pi}\sum_{\alpha=1}^{m_i}\phi_{\alpha}(0,t)-{\cal N}_i(V^{(i)}_{g})\right]^{2}.\;\;\;\;
\label{sc}
\end{eqnarray} 
accounts for the mesoscopic Coulomb blockade \cite{AleiGlaz} in the metallic left/right QDs characterized by the charging energies $E^{(i)}_C$.  Here ${\cal N}_i(V^{(i)}_g)$ are the  dimensionless parameters controlled by the gate voltages $V_g^{(i)}$ and $n_\tau(t)=\theta(t)\theta(\tau-t)$ is a function counting the number of electrons entering the QDs area,  $\theta(t)$ is a Heaviside (step) function.

The backscattering action in the left/right parts of the circuit is given by the boundary sine-Gordon model:
\begin{eqnarray}
S^{(i)}_{\rm bs} & = & -\frac{D}{\pi}\;\sum_{\alpha=1}^{m_i} {\color{black}|r_\alpha|}\int_{0}^{\beta}dt \cos\left[2\phi_{\alpha}(0,t)\right]\;\;\;
\label{bs}
\end{eqnarray}
Here $|r_\alpha|$ are reflection amplitudes of $\alpha$-QPCs,  $D$ is a bandwidth (ultraviolet cutoff of the theory). As the QPCs do not talk to each other, we introduce an independent one-dimensional coordinate systems ($x_\alpha$ axes) for each QPC separately  (see Fig. \ref{f1}).

Two circuits are connected through the tunnel contact (dashed lines in the center of Fig.\ref{f1}).  Corresponding tunnel action is given by:
\begin{equation}
S^{(12)}_{\rm tun}= -\int_0^\beta d t \left[ t_{12} \bar \Psi_1(-\infty, t) \Psi_2(-\infty, t) +h.c.\right]\;\;\;
\end{equation}
The operators $\Psi_i (x=-\infty, t)$ denote the fermions in the QD$_i$ at the position of the left/right side of the tunnel contact.

{\color{blue} \textit{Transport integrals}} - The charge current 
$I_e$ and heat current $I_h$ depend on the temperature drop $\Delta T$ and voltage drop $\Delta V$  across the tunnel barrier \cite{book1, T_Review}.  Assuming the linear response if both 
$[\Delta T,\Delta V]\ll T$ we define coupled transport equations
\begin{equation}
 \left(%
\begin{array}{c}
  I_{e} \\
  I_{h} \\
\end{array}%
\right)= \left(%
\begin{array}{cc}
  L_{11} & L_{12} \\
  L_{21} & L_{22} \\
\end{array}%
\right)\left(%
\begin{array}{c}
  \Delta V \\
  \Delta T \\
\end{array}%
\right).
\label{kk}
\end{equation}
The diagonal coefficients of the matrix ${\bf L}$  are defined as 
\begin{equation}
G=L_{11}=\left.\frac{\partial I_e}{\partial \Delta V}\right|_{\Delta T=0},\;\;\;\;\;
G_H=L_{22}=\left.\frac{\partial I_h}{\partial \Delta T}\right|_{\Delta V=0}\;\;\;\;
\end{equation}
and the off-diagonal coefficients are given by
\begin{equation}
G_T=L_{12}=\left.\frac{\partial I_e}{\partial \Delta T}\right|_{\Delta V=0}=\frac{1}{T}\left.\frac{\partial I_h}{\partial \Delta V}\right|_{\Delta T=0}=\frac{L_{21}}{T}\;\;\;\;
\end{equation}
The thermo-electric power (thermopower, Seebeck coefficient) ${\cal S}$ and thermal conductance ${\cal K}$ are defined at zero-electric-current state $I_e=0$ as
\begin{equation}
{\cal S}=-\left.\frac{\Delta V}{\Delta T}\right|_{I_e=0}=\frac{G_T}{G}
\end{equation}
and
\begin{equation}
{\cal K} = \left.\frac{\partial I_h}{\partial \Delta T}\right|_{I_e=0}=\frac{{\rm det} {\bf L}}{L_{11}} = G\cdot T\left[\frac{G_H}{G\cdot T} - {\cal S}^2\right]\;\;\;\;
\end{equation}
The Wiedemann-Franz (WF) law  \cite{book1, T_Review} establishes a connection between the thermal conductance ${\cal K}$ and electrical conductance $G$  through an universal constant, the Lorenz number ${\cal K}/(GT)$$=$$L_0$$=$$\pi^2/3$.  
The validity of WF law is attributed to the fact that both charge and heat are transferred by the same quasiparticles.
The deviation from the relation  ${\cal K}/(L_0GT)$$=$$1$ is sometimes called "violation of the WF law". In this {\it Letter} we show that the WF law can be understood in more general terms while the fundamental constant is not necessarily coincides with $L_0$.

To proceed with calculation of the charge and heat transport through the two-site Kondo circuit we define transport integrals (see \cite{T_Review,nk2022}):
\begin{equation}
{\cal L}_n(T) = \frac{1}{4T} \int_{-\infty}^{+\infty} d \epsilon \frac{\epsilon^n}{\cosh^2\left(\epsilon/2T\right)} {\cal T}(T,\epsilon),\;\;\;n=0,1,2.
\label{ti}
\end{equation}
where we denote by $ {\cal T}(T,\epsilon)$ a transmission coefficient
\begin{equation}
 {\cal T}(T,\epsilon)=2\pi |t_{12}|^2\nu_1(\epsilon, T) \nu_2(\epsilon, T).
 \label{transco}
\end{equation}
Here  the local densities of state (DoS) $\nu_i(\epsilon,T)$ at the position of the tunnel barrier  are given by 
\begin{eqnarray}
\nu_i(\epsilon,T)&=& -\frac{1}{\pi}\cosh\left(\frac{\epsilon}{2T}\right)
\int_{-\infty}^\infty {\cal G}_i\left(\displaystyle\frac{1}{2T} +it\right)
e^{i\epsilon t} d t,\;\;\;\;\;\;\;\;
\label{dos}
\end{eqnarray}
The DoS are defined in terms of electron's Green's functions ${\cal G}_i(\tau)=-\langle T_\tau \Psi_i(-\infty,\tau)\bar\Psi_i (-\infty,0)\rangle$ where  $T_\tau$ is the imaginary time-ordering
\begin{equation}
{\cal G}_i(\tau)=-\frac{\nu^{(i)}_{0}\pi T}{\sin\left(\pi T \tau\right)}K_i(\tau)
\end{equation}
and $\nu^{(i)}_0$ are bare (non-renormalized)  DoS in QD$_i$.
The correlators $K_i(\tau)$ account for the effects of interaction in the left/right
parts of the Kondo circuit and are computed e.g. by using the bosonization technique \cite{GNTBook}.
The connections between the transport integrals (\ref{ti}) and kinetic coefficients (\ref{kk}) are as follows: $L_{11}=  {\cal L}_0$, $L_{12}= - {\cal L}_1/T$ and  $L_{22} ={\cal L}_2/T$.

Lorenz ratio $R_L(T,{\cal N}_1,{\cal N}_2)$ has the following definition  in terms of the transport integrals:
\begin{equation}
R_L(T,{\cal N}_1,{\cal N}_2)= \frac{3}{(\pi T)^2}\left[\frac{{\cal L}_2}{{\cal L}_0} - \left(\frac{{\cal L}_1}{{\cal L}_0} \right)^2\right]
\end{equation}
There are two contributions to $R_L(T,{\cal N}_1,{\cal N}_2)$ which behave differently at low and high temperatures
\begin{equation}
R_L(T,{\cal N}_1,{\cal N}_2)= {\cal R}(T,{\cal N}_1,{\cal N}_2)- \frac{3}{\pi^2} {\cal S}^2(T,{\cal N}_1,{\cal N}_2).
\label{lora}
\end{equation}
One is ${\cal R}(T,{\cal N}_1,{\cal N}_2)= 3/(\pi T)^2({\cal L}_2/{\cal L}_0)$ and another one is proportional to the square of the thermopower $S(T,{\cal N}_1,{\cal N}_2)={\cal L}_1/(T\cdot {\cal L}_0)$. Both contributions  depend on the temperature $T$ and the dimensionless gate voltages ${\cal N}_i$.  The Wiedemann-Franz law constitutes $R_L={\cal R}=1$ at all temperatures and all gate voltages.  Strictly speaking, this law is not satisfied exactly at any given set of parameters and therefore is always violated. However, we can adopt some more general definition of the WF law, for KC,  namely
{\it if there exists some parametric region of the temperatures and gate voltages
at which the main contribution to $R_L$ is given by an universal constant
and the non-universal corrections to it are controllably and  vanishingly small},
we conclude that the generalized WF law is satisfied.  The question of
whether the generalized WF law is violated or not is therefore reformulated as a problem of computing $R_L$ and finding out whether or not it acquires some non-trivial value.
Besides, if this non-trivial value is different from unity,  it is interesting and important to know what kind of useful information the WF law conveys.

Let us first summarize the key equations for the transport integrals in terms of the 
correlators $K_i$ \cite{thanh2018}:
\begin{equation}
{\cal L}_0(T)= \frac{g_C}{2}\int_{-\infty}^{\infty} \frac{d z }{\cosh^{2}z} \displaystyle K_{1}^+(z,T)\cdot K_{2}^-(z,T)
\end{equation}
where $g_C=2\pi \nu^{(1)}_{0}\nu^{(2)}_{0}|t_{12}|^{2}$ is a conductance of the central tunnel area. We denote kernels $K_{i}^\pm (z, T) = K_{i}\left((\pi/2 \pm  i z )/(\pi T)\right)$
obeying obvious symmetry property
$K_{i}^\pm (-z, T) = K_{i}^\mp (z, T) $.  Here $z=\pi T t$ is a dimensionless time.  We explicitly assume an additional temperature dependence of the pre-factors of the kernels $K_i$ (see discussion below).

The equation for ${\cal L}_1(T)$  is written  as follows \cite{thanh2018}
\begin{eqnarray}
{\cal L}_1= i (\pi T) \frac{g_C}{4}
\int_{-\infty}^{\infty}\frac{d z}{\cosh^{2} z}
W[ K_1^+(z,T), K_2^-(z,T)],\;\;\;\;\;\;\;\;
\label{cond3c}
\end{eqnarray}
in terms of the Wronskian of two kernels:
\begin{eqnarray}
W\left[K_1^+\;
K_2^-\right]=
\left|
\begin{array}{cc}
\displaystyle
K_1^+(z,T)  & \displaystyle K_2^-(z,T)\\
\partial_z K_1^+(z,T) & \partial_z K_2^-(z,T)
\end{array}
\right|
\label{wronskian}
\end{eqnarray}
{\color{black} If the particle-hole  (PH) symmetry in two-site Kondo circuit holds,  both kernels $K_1^\pm$ and $K_2^\mp$ are even functions of $z$ (for a symmetric Kondo circuit kernels are linear dependent). } As a result, both the ${\cal L}_1$ coefficient and thermopower vanish.  However, the backscattering (\ref{bs}) 
breaks the PH symmetry
and therefore leads to the finite value (at finite temperature and certain parametric region of the gate voltages) of the Seebeck coefficient ${\cal S}$. We note,
that the smallness of this coefficient is controlled by the smallness of the PH symmetry breaking parameter.  The PH symmetry is protected both at Coulomb valleys {\color{black} (integer ${\cal N}$)} and Coulomb peaks 
{\color{black} (half-integer ${\cal N}$)} where thermopower is exactly zero.
In addition, the thermopower vanishes at low temperature regime \cite{T_Review}. Indeed, exact calculations for the $m_2$$=$$M$$=$$1$ channel charge Kondo circuit setup in contact with the normal metal ($m_1$$=$$N$$=$$\infty$)  predict \cite{andreevmatveev}
for the amplitude of the Seebeck coefficient oscillations 
${\cal S}_{\rm max}$$ \propto $$T/E^{(2)}_C$ while for $M=2$ thermopower ${\cal S}_{\rm max} $$\propto $$ \sqrt{T/E^{(2)}_C}\ln (E^{(2)}_C/T)$ \cite{andreevmatveev} (see the Table \ref{tab1}).  The perturbative (in terms of the small backscattering amplitudes 
$|r_\alpha|$$\ll $$ 1$) calculations for the $M>2$ ,  $N=\infty$ Kondo circuits results in ${\cal S}_{\rm max}$$ \propto $$\sqrt[3]{T/E^{(2)}_C}\ln (E^{(2)}_C/T)$ \cite{thanhprl}.  We therefore conclude that the thermopower contribution to the Lorenz ratio $R$ vanishes at sufficiently low temperatures $T\ll E^{(i)}_C$ {\color{black} independently on imposing the particle-hole symmetry} and therefore can be disregarded at that limit.  It is sufficient for the verification of the WF law to {\it  compute} the values of both ${\cal L}_0$ and  ${\cal L}_2$ at the PH-symmetric point assuming all 
$|r_\alpha|=|r|=0$. Since the amplitude of the mesoscopic Coulomb blockade oscillations is proportional to $|r|$ \cite{AleiGlaz}, the limit $|r|\to 0$ washes out completely the 
${\cal N}_i$ dependence of the Lorenz ratio.   {\color{black} It is sufficient therefore to {\it measure} the electric and  the thermal conductances at $T\ll E^{(i)}_C$ close to Coulomb peaks \cite{com1} to verify predicted magic Lorenz ratios.}

The ${\cal L}_2$ transport coefficient is written in terms of dimensionless time integrals as follows \cite{nk2022}:
\begin{widetext}
\begin{equation}
{\cal L}_2(T)= (\pi  T)^2\cdot \frac{g_C}{2}
\int_{-\infty}^{\infty} d z
\frac{\left(2-\cosh^2 [z]\right)\cdot  K_{1}^+(z,T)\cdot
K_{2}^-(z,T) +\cosh^2 [z]\cdot \partial_z K_{1}^+(z,T)\cdot
\partial_z  K_{2}^-(z,T)}{\cosh^4 [z]}
\end{equation}
\end{widetext}
\noindent
{\color{blue} \textit{Results and discussion}}
- As only the charge mode $\phi_c(0,t)=1/\sqrt{m_i} \sum_{\alpha=1}^{m_i}\phi_\alpha(0,t)$ enters  the Coulomb blockade action (\ref{sc}), the PH symmetric part of the kernels $K_i$ can be obtained from the $m_i=1$  result (see \cite{furusakimatveevprb}) by doing a simple rescaling $E^{(i)}_C$$\to$$m_i$$\cdot$$E^{(i)}_C$, $n_\tau$$\to$$n_\tau/\sqrt{m_i}$ and ${\cal N}_i$$\to$${\cal N}_i$$/\sqrt{m_i}$ (see details of derivation in \cite{andreevmatveev} and also in Supplemental Materials of Ref.  \cite{thanhprl}). 
Evaluating the Gaussian action $S^{(i)}_0+S^{(i)}_C$ (\ref{act0}),(\ref{sc}) with the saddle point method \cite{andreevmatveev} and computing the fluctuations around the saddle point similarly to \cite{andreevmatveev}  we obtain:
\begin{eqnarray}
&&\ln K_i^\pm(\tau) \left|_{r=0}\right.=-2E^{(i)}_C T\sum_{\omega_n}
\frac{[1-\cos\omega_n \tau]e^{-|\omega_n|/D}}{|\omega_n|\left(|\omega_n|+m_i E^{(i)}_C/\pi\right)}\nonumber\\
&&\approx \frac{2}{m_i}\ln\left(\frac{\pi^2 T}{m_i \gamma E^{(i)}_C\left|\sin[\pi T\tau]\right|}\right),
\end{eqnarray}
Here we performed a summation over bosonic Matsubara frequencies $\omega_n=2\pi T n$ assuming the limit $\tau\gg [E^{(i)}_C]^{-1}$.  Details for similar calculations of Matsubara sums can be found in \cite{furusakimatveevprb, andreevmatveev, thanhprl}.
Applying simultaneously a shift transformation and a Wick rotation from imaginary to real time 
$\tau \to \frac{1}{2T}+it$ we finally get:
\begin{equation}
K_i ^\pm(z,T)\left|_{r=0}\right. =\frac{A_i(T)}{\cosh^{2/m_i}[z]},\;
A_i(T)=\left(\frac{\pi^2 T}{\gamma E^{(i)}_C  m_i}\right)^{2/m_i}.
\label{kernel}
\end{equation}
Here $\gamma $$=$$ e^C$ and $C$$=$$0.577$ is the Euler's constant.
The Eq.(\ref{kernel}) is the central point for the calculation of the Lorenz ratio. In particular,  the power law behaviour of the kernel leads to some particular temperature behaviour of the electric conductance attributed to the Anderson's orthogonality catastrophe.  For example, if $m_1$$=$$\infty$ and $m_2$$=$$M$ the conductance scales as 
$G$$\propto $$T^{2/M}$.  The explanation of this behaviour for a particular case $M$$=$$2$ and it's connection to the Anderson's orthogonality catastrophe was given in a seminal Matveev-Furusaki (MF) paper \cite{furusakimatveevprb}. Assuming that the left and the right parts of the two-site charge Kondo circuit are separated by the tunnel barrier and therefore can be treated independently as an electron loses its coherence, we sketch the MF arguments (for the sake of the Reader's convenience) for arbitrary value of Kondo channel's number $M$$\geq $$2$.  {\it As the charge fluctuations are not suppressed below the energies $E^{(2)}_C$, one can interpret the effects of charging energy as a hard-wall boundary condition for the wave function. When the electron tunnels through the barrier from the left part of the Kondo circuit (KC) (let us 
call the left part of the circuit a "lead" for the right part of KC), the charging energy of the KC is lowered by moving one electron through the right part containing $M$ identical QPCs. Therefore, each mode (QPC) transfers $q$$=$$\pm e/M$ charge (we consider both electron's and hole's transport). The Friedel's sum rule tells that the corresponding phase shifts are 
$\delta$$ =$$\pm\pi/M$. Sudden change of the boundary condition is accompanied by  the large number of the electron-hole pair excitations and results in a creation of  a new state which is almost orthogonal to the ground state of the system. The orthogonality leads to
a suppression of the tunnelling density of states (\ref{dos}) 
$\nu (\epsilon)$$\propto $$\epsilon^\chi$
where according to Friedel's sum rule $\chi$$=$$\sum(\delta/\pi)^2$ and sum is taken over all modes. The total number of the modes is $n=2M$ ($M$ modes in the dot and $M$ modes in the lead). As a result, $\chi$$=$$2 M/M^2$$=$$2/M$ and therefore 
$\nu_2$$\propto $$\epsilon^{2/M}$ which leads to corresponding  temperature scaling of the transport coefficient ${\cal L}_0$ \cite{furusakimatveevprb}}. 
The temperature scaling of ${\cal L}_1$ is determined by the transport integral containing 
$\epsilon\cdot\nu_2(\epsilon)$$\propto $$ \epsilon^{1+2/M}$ (\ref{ti}) and for 
${\cal L}_2$,  corresponding equation  (\ref{ti})  contains 
$\epsilon^2\cdot\nu_2(\epsilon)$$\propto $$\epsilon^{2+2/M}$. The same arguments can be repeated for the left part of the KC containing  $N$ independent QPCs and treating the right part as a contact. The OC results in $\nu_1$$\propto $$\epsilon^{2/N}$.

To  compute the ratio ${\cal L}_2/{\cal L}_0$ we ignore
the exact form of the temperature dependent pre-factor in $K_i$ (\ref{kernel}) which will be cancelled out  and obtain for ${\cal L}_0$:
\begin{eqnarray}
&&{\cal L}_0=A_1\cdot A_2 \cdot  \frac{g_C}{2} \int_{-\infty }^{\infty } \frac{d z}{\cosh ^{2+2/N+2/M}[z]}\\
&&=\left(\frac{\pi^2 T}{\gamma E^{(i)}_C  N}\right)^{2/N}
\left(\frac{\pi^2 T}{\gamma E^{(i)}_C  M}\right)^{2/M}
\frac{g_C}{2} \frac{\sqrt{\pi } \Gamma \left(1+\frac{1}{N}+\frac{1}{M}\right)}{\Gamma \left(\frac{3}{2}+\frac{1}{N}+\frac{1}{M}\right)}\nonumber
\label{L0}
\end{eqnarray}
Similar procedure is applied  for  the calculation of ${\cal L}_2$:
\begin{eqnarray}
&&\frac{{\cal L}_2}{(\pi  T)^2}=A_1\cdot A_2  \cdot \frac{g_C}{2} \int_{-\infty }^{\infty } \frac{\left(2-\cosh ^2 [z]+\frac{4\sinh^2 [z]}{NM}\right)}{\cosh^{4+2/N+2/M}[z]} \, dz\nonumber\\
&&=\left(\frac{\pi^2 T}{\gamma E^{(i)}_C  N}\right)^{2/N}
\left(\frac{\pi^2 T}{\gamma E^{(i)}_C  M}\right)^{2/M}\frac{g_C}{2}\times\nonumber\\
&&\times  \frac{\sqrt{\pi } (M+2) (N+2) \Gamma \left(1+\frac{1}{N}+\frac{1}{M}\right)}{2 M N \Gamma \left(\frac{5}{2}+\frac{1}{N}+\frac{1}{M}\right)}
\label{L2}
\end{eqnarray}
Here $\Gamma(z)$ is Euler's gamma function.
Substituting these function to equation (\ref{lora}), omitting vanishing at  $T\ll E_C^{(i)}$ term  ${\cal S}^2$ and disregarding weak non-universal gate-voltage-dependent corrections we finally get $R_L$$($$T$$\to$$ 0$$)$$=$${\cal R}_{N,M}$, where
\begin{equation}
{\cal R}_{N,M}=
\frac{3 (M+2) (N+2)}{3NM +2N+2 M}
\label{lora_ex}
\end{equation}
It directly follows from  (\ref{lora_ex})
that the maximal value of ${\cal R}_{\rm max}=27/7$ is achieved at $N=M=1$ when the orthogonality catastrophe leads to the maximal suppression of the density of states.
This value is quite close to the absolute upper bound ${\cal R}_{\rm ub}=21/5$
obtained in the work \cite{karki2}.  {\color{black} The upper bound  \cite{karki2} is obtained assuming that  {\it the system
modelled by the
scattering theory and the transmission coefficient is merely
energy dependent, the temperature comes solely from
the Fermi-function \cite{karki2}}. The orthogonality catastrophe results in a specific temperature scaling of the transmission coefficient (\ref{transco})
${\cal T}$$($$T$$,$$x$$=$$\epsilon/$$T$$)$$\left.\right|_{r=0}$$\propto$$T^{\frac{2}{N}+\frac{2}{M}}$$ f_{NM}(x)$ which vanishes 
at $T\to 0$ limit.  Here $f_{NM}(x)$ is some function depending on the number of channels and dimensionless energy $x$ \cite{com2}. Therefore, the upper bound ${\cal R}_{\rm max}=27/7$ represents the maximal value of the Lorenz ratio for the non-Fermi liquid transmission coefficient. }
Interestingly, the Lorenz ratio provides unique benchmark for the orthogonality catastrophe.  The values of ${\cal R}$ are  different even when the temperature dependence of 
${\cal L}_0$ and ${\cal L}_2$ are the same for different two-site Kondo circuits (compare, e.g $N=1$,  $M=\infty$ and $N=M=2$,  see Table \ref{tab1}).

Expanding the general equation for ${\cal R}$ (\ref{lora_ex}) for the large values of $N$ and $M$ 
we conclude  that the Lorenz ratio ${\cal R}$  is bounded from below by its minimal value ${\cal R}_{\rm min}=1$ constituting the conventional Wiedemann-Franz law.
Is ${\cal R}$ {\it always} different from unity in the strongly correlated systems? In fact, not. We discussed the behaviour of transport integrals in the simulator where both electric conductance $G$ and ratio ${\cal K}/T$ vanish at low temperatures. 
We argue that in contrast to the statement of \cite{sela}, the value of Lorenz ratio ${\cal R}$ is universal due to the orthogonality catastrophe despite of vanishing $G$ and ${\cal K}/T$.  If, however, both 
quantities remain finite at the $T\to 0$ limit (which can be true both in Fermi- and 
Non- Fermi liquid regimes, see \cite{karkis, sela}), the WF law is satisfied for the non-interacting leads with the Lorenz ratio
${\cal R}=1$ \cite{karkis, sela}.   {\color{black} The power-law or logarithmic temperature non- analyticity of the transmission coefficient close to a critical non-Fermi liquid intermediate coupling fixed point \cite{karkis} only results in vanishing at low temperature/energy corrections to the Lorenz ratio. }  It is worth mentioning that the effects of interaction in quantum wires (leads) attached to the nano-devices (e.g. mesoscopic islands hosting local modes \cite{nava1, nava2}) results in a deviation from conventional WF law. In that case the proportionality coefficient (Lorenz ratio) provides some important information about the effects of interaction in the quantum wires.  

\begin{table}
\begin{ruledtabular}
\begin{tabular}{|| p{0.08\linewidth}  | p{0.08\linewidth}  |   p{0.1\linewidth}  |   p{0.1\linewidth}   | p{0.1\linewidth}  | p{0.18\linewidth}  ||} 
 \hline
N & M &  ${\cal R}$ & ${\cal L}_0$ & ${\cal L}_2$ & ${\cal S}_{\rm max}$
\\ [0.5ex] 
 \hline\hline
 1 & 1 & 27/7 &$T^4$ & $T^6$ & $T$\\ 
 \hline
  1 & 2 & 3 & $T^3$  & $T^5$ & $\sqrt{T}\ln T$\\ 
 \hline
 1 & 3 & 45/17  &$T^{8/3}$ & $T^{14/3}$ & $\sqrt[3]{T}\ln T$ $^\blacklozenge$\\ 
 \hline
 1 & $\infty$  & 9/5$^\bigstar$& $T^2$ & $T^4$ & $T$\\ 
 \hline
 2 & 2 & 12/5 & $T^2$ & $T^4$ & $T$\\ 
 \hline
 2 & 3 & 15/7 & $T^{5/3}$ & $T^{11/3}$ & $\sqrt[3]{T}\ln T$ $^\blacklozenge$\\ 
 \hline
 2 & $\infty$ & 3/2$^\bigstar$ & $T$ & $T^3$ & $\sqrt{T}\ln T$\\ 
 \hline
 3 & 3 & 25/13 & $T^{4/3}$ & $T^{10/3}$& $\sqrt[3]{T}\ln T$ $^\blacklozenge$\\ 
 \hline
 3 & $\infty$ & 15/11 & $T^{2/3}$ & $T^{8/3}$ & $\sqrt[3]{T}\ln T$ $^\blacklozenge$\\ 
 \hline
$ \infty$ & $\infty$ & 1 & $T^0$ & $T^2$ &0 \\ 
 \hline
\end{tabular}
\end{ruledtabular}
\caption{Magic Lorenz ratios ${\cal R}$ for the two-site Kondo circuit connecting $N$- and $M$- channel Kondo simulators operating in either Fermi or Non-Fermi liquid regimes
from Eq. (\ref{lora_ex}). The last three columns show the temperature dependence of the diagonal transport integrals 
${\cal L}_0$$\propto $$A_1$$\cdot $$A_2$ and 
${\cal L}_2$$\propto $$A_1$$\cdot $$A_2$$\cdot $$T^2$ 
(see Eqs. (\ref{kernel} - \ref{L2}))
and thermopower at low temperatures.  Results marked by $^\bigstar$ were reported in \cite{karki1}.  Mark $^\blacklozenge$ refers to the perturbative results \cite{thanhprl}.}
\label{tab1}
\end{table}
{\color{blue} \textit{Conclusions}} - Summarizing, we checked the validity of the Wiedemann-Franz law in the two-site Kondo circuits.  The circuits consist of two Kondo simulators 
operating either in strongly correlated Fermi- or the Non-Fermi liquid regimes. The two parts of the circuit are  connected by the tunnel contact. It is shown that the proportionality between thermal and charge conductances holds even for the case of strong electron-electron correlations.  The transport integrals satisfy the generalized Wiedemann-Franz law at low temperatures with the magic Lorenz ratios which are always greater than one. The magic Lorenz ratios contain some important information about the Anderson's orthogonality catastrophe and provide a number benchmark for the unique characterization of the two-site Kondo circuit operational regime. The "two-islands" experimental setups \cite{two_islands} can be directly used for verification of 
the generalized WF law and OC predictions.

{\color{blue} \textit{Acknowledgements}} - M.N.K is thankful to Thanh Nguyen and Deepak Karki for numerous fruitful discussions of the thermoelectric transport through the Kondo simulators. This work is conducted within the framework of the Trieste Institute for Theoretical Quantum Technologies (TQT).  M.N.K acknowledges the support from the Alexander von Humboldt Foundation for the research visit to IFW Dresden.


\vspace*{-3mm}
\begin{thebibliography}{10}
\vspace*{-3mm}

\bibitem{orthogonal} P.W. Anderson, Phys. Rev. Lett {\bf 18}, 1049 (1967).

\bibitem{Mahan} G.D. Mahan, Phys. Rev. {\bf 163}, 612 (1967).

\bibitem{Noz1} B.Roulet,  J.Gavoret and P.Nozieres, Phys. Rev. {\bf 178}, 1072 (1969).

\bibitem{Noz2} P.Nozieres, J.Gavoret, B.Roulet,  Phys. Rev. {\bf 178}, 1084 (1969).

\bibitem{Noz3} P.Nozieres and C.T De Dominicis,  Phys. Rev. {\bf 178}, 1097 (1969).

\bibitem{Kondo} J. Kondo, Prog. Theor. Phys. {\bf 32}, 37 (1964).

\bibitem{Hewson} A. C. Hewson, {\it The Kondo Problem to Heavy Fermions}, Cambridge University Press, Cambridge, 1993.

\bibitem{speed} Th. Fogarty,  S. Deffner, Th. Busch,  and S. Campbell, Phys. Rev. Lett. {\bf 124}, 110601 (2020).

\bibitem{OC_UG} M. Knap,  A. Shashi, Y. Nishida, A. Imambekov, D. A. Abanin, and E. Demler, Phys. Rev. {\bf X 2},  041020 (2012).

\bibitem{NB} Ph. Nozieres and A. Blandin,  J. Physique {\bf 41}, 193 (1980).

\bibitem{thanh2018} T. K. T. Nguyen, M. N. Kiselev, Phys. Rev. B
\textbf{97}, 085403 (2018).

\bibitem{pierre2} Z. Iftikhar, S. Jezouin, A. Anthore, U. Gennser,
F. D. Parmentier, A. Cavanna and F. Pierre, Nature \textbf{526},233(2015).

\bibitem{pierre3} Z. Iftikhar, A. Anthore, A. K. Mitchell, F. D.
Parmentier, U. Gennser, A. Ouerghi, A. Cavanna, C. Mora, P. Simon,
and F. Pierre, Science \textbf{360},1315 (2018).

\bibitem{Flensberg} K. Flensberg, Phys. Rev. B \textbf{48},11156 (1993).

\bibitem{Matveev} K. A. Matveev, Phys. Rev. B \textbf{51},1743 (1995).

\bibitem{furusakimatveevprb} A. Furusaki, K. A. Matveev, Phys. Rev.
B \textbf{52}, 16676 (1995).

\bibitem{AleiGlaz} I. L. Aleiner and L. I. Glazman, Phys. Rev. B {\bf 57}, 9608 (1998).

\bibitem{two_islands} W. Pouse, L. Peeters, C. L. Hsueh, U. Gennser,
A. Cavanna, M. A. Kastner, A. K. Mitchell, D. Goldhaber-Gordon,  Nature Phys. (2023). {\rm https://doi.org/10.1038/s41567-022-01905-4}

\bibitem{Karki2022a} D. B. Karki, E. Boulat, and C. Mora,  Phys. Rev. B {\bf 105}, 245418  (2022).

\bibitem{Karki2023}  D. B. Karki, E. Boulat, Winston Pouse,  D.  Goldhaber-Gordon, 
A. K. Mitchell, C. Mora,  Phys. Rev. Lett {\bf 130}, 146201 (2023).

\bibitem{thanhprl} T. K. T. Nguyen, M. N. Kiselev, Phys. Rev. Lett.
\textbf{125}. 026801 (2020).

\bibitem{Sela1}  C.  Piquard, P.  Glidic, C. Han, A. Aassime, A. Cavanna, U. Gennser, Y. Meir, E. Sela, A. Anthore, F. Pierre,  arXiv:2303.12039 (2023).

\bibitem{Sela2}  E. Sela, D. Goldhaber-Gordon, A. Anthore,  F.  Pierre, Y. Oreg, arXiv:2302.02295 (2023)

\bibitem{andreevmatveev} A. V. Andreev, K.A. Matveev,  Phys. Rev.
Lett. \textbf{86}, 280 (2001); K.A. Matveev, A. V. Andreev,  Phys. Rev. B \textbf{66},045301 (2002) .


\bibitem{LeHur2002} K. Le Hur and G. Seelig, Phys. Rev. B \textbf{65}, 165338
(2002).

\bibitem{thanh2010} T. K. T. Nguyen, M. N. Kiselev, and V. E. Kravtsov,
Phys. Rev. B,  \textbf{82}, 113306 (2010).


\bibitem{GNTBook} A. O. Gogolin, A. A. Nersesyan, and A. M. Tsvelik, 
{\it Bosonization Approach to Strongly Correlated Systems} (Cambridge
University Press, Cambridge, UK, 1998.

\bibitem{comment1} For simplicity we consider non-interacting 1D fermions
describing IQH $\nu=1$ chiral edge mode. The effects of interactions in the Luttinger Liquid formed at a narrow constriction (QPC)  was recently considered in \cite{int1,int2,int3}.

\bibitem{int1} T.K.T. Nguyen and M.N. Kiselev,
Comm.  in Phys. , {\bf 30},  1 (2020)

\bibitem{int2}  A.V. Parafilo, T.K.T. Nguyen and M.N. Kiselev,
Phys. Rev.  B {\bf 105}, L121405 (2022).

\bibitem{int3}  T. K. T. Nguyen, A. V. Parafilo, H. Q. Nguyen,and M. N. Kiselev,
Phys. Rev.  B {\bf 107},  L201402 (2023).

\bibitem{book1}V. Zlatic and R. Monnier, {\it Modern Theory of Thermoelectricity}, Oxford University Press,  2014.

\bibitem{T_Review} G. Benenti, G. Casati, K. Saito, and R. Whitney, Phys. Rep. 694,  1 (2017).

\bibitem{com1}{\color{black} The orthogonality catastrophe develops in the vicinity of the Coulomb peaks (half- integer ${\cal N}$) \cite{furusakimatveevprb} 
where the Majorana's level width 
$\Gamma({\cal N})$$\propto $$ |r|^2 $$E_C $$\cos^2(\pi {\cal N})$$\to $$0$ \cite{furusakimatveevprb}. Away from the Coulomb peaks the condition $\Gamma ({\cal N}) $$ \ll $$ T$$ \ll  $$ E_C$ should be applied for the experimental verification of magic Lorenz ratios \cite{furusakimatveevprb}.}

\bibitem{nk2022}  T.K.T. Nguyen and M.N. Kiselev,
Comm. in Phys. {\bf 32},  331 (2022).

\bibitem{karki2} D. B. Karki, Phys. Rev. B \textbf{102}, 115423 (2020).

\bibitem{com2} {\color{black} Asymptotic of $f_{NM}(x)$$=$${\rm const}(N,M) $$+$$ O(x^2)$ for $x$$\ll $$1$ can be obtained by Taylor expansion in the equation (\ref{dos}).  The asymptotic of transmission coefficient at $\epsilon$$\gg $$ T$ 
(or, equivalently, $x$$\gg $$1$) is  
${\cal T} (T,\epsilon)\left.\right|_{r=0} $$\propto $$ \epsilon^{2/N+2/M}$ due to  the orthogonality catastrophe (see discussion in the main text).}

\bibitem{karki1} D. B. Karki, Phys. Rev. B \textbf{102}, 245430 (2020).


\bibitem{sela} G. A. R. van Dalum, A. K. Mitchell, and L. Fritz, Phys. Rev.
B {\bf 102}, 041111(R) (2020).

\bibitem{karkis}  D.B. Karki and M.N. Kiselev,  Phys.  Rev. B {\bf 102}, 241402(R) (2020).

\bibitem{nava1} D. Giuliano, A. Nava, R. Egger, P. Sodano, and F. Buccheri,  Phys. Rev.  B {\bf 105}, 035419 (2022).

\bibitem{nava2} F.  Buccheri, A. Nava, R. Egger, P. Sodano, and D. Giuliano,  Phys. Rev.  B {\bf 105}, L081403 (2022).

\end{thebibliography}
\end{document}